\documentclass[twocolumn,aps,prb,groupedaddress]{revtex4-1}
\usepackage{graphicx}
\usepackage{color}
\usepackage{amsmath}

\newcommand{\be}{\begin{equation}}
\newcommand{\ee}{\end{equation}}
\newcommand{\bea}{\begin{eqnarray}}
\newcommand{\eea}{\end{eqnarray}}

\newcommand{\la}{\langle}
\newcommand{\ra}{\rangle}
\newcommand{\lb}{\left[}
\newcommand{\rb}{\right]}
\newcommand{\lp}{\left(}
\newcommand{\rp}{\right)}

\renewcommand{\vec}[1]{{\bf #1}}

\begin{document}
\title{Energy Flows in Graphene: Hot Carrier Dynamics and Cooling}

\author{Justin C. W. Song$^{1,2}$}
\author{Leonid S. Levitov$^1$}
\affiliation{$^1$ Department of Physics, Massachusetts Institute of Technology, Cambridge, Massachusetts 02139, USA}
\affiliation{$^2$ School of Engineering and Applied Sciences, Harvard University, Cambridge, Massachusetts 02138, USA}

\begin{abstract}
Long lifetimes of hot carriers 
can lead to qualitatively new types of responses in materials. The magnitude and time scales for these responses reflect the mechanisms governing energy flows. We examine the microscopics of two processes which are key for energy transport, focusing on the unusual behavior arising due to graphene's unique combination of material properties. One is hot carrier generation in its photoexcitation dynamics, where hot carriers multiply 
through an Auger type carrier-carrier scattering  cascade. The hot-carrier generation manifests itself through elevated electronic temperatures which can be accessed 
in a variety of ways, in particular optical conductivity measurements. Another process of high interest is electron-lattice cooling. We survey different cooling pathways and discuss the cooling bottleneck arising for the momentum-conserving electron-phonon scattering pathway. We show how this bottleneck can be relieved by higher-order collisions - supercollisions - and examine the variety of supercollision processes that can occur in graphene.
\end{abstract} 
\maketitle

\section{Introduction}

Every so often we encounter things that neatly combine beauty and utility. In a
fairy tale a  much-admired flower---``the sacred lotus of Hindostan''---turns
out to be a common flower from the kitchen-garden, an artichoke.\cite{andersen}
Likewise, the subject of energy flows in materials
is aesthetically pleasing and, at the same time, harbors practical
opportunities.
%
Understanding energy transport 
mechanisms 
is of keen interest for designing new approaches to handle, convert, and utilize energy in a bid to 
address key technological challenges. 
One area of high current interest which may benefit from this research is IT hardware, in particular finding ways to circumvent the saturation of operating frequencies in integrated electronics due to the large amounts of power dissipated in microprocessors.\cite{pop10} Another such area is the development of efficient solar cells, where the relation with hot carriers stems from the Shockley-Queisser limit that sets an upper bound for conversion efficiencies in single-junction solar cells.\cite{sqlimit} Currently high expectations are pinned on  
two-dimensional (2D) materials,\cite{vanderwaals} such as graphene and the atomically thin dichalcogenides, 
which possess a number of potentially useful properties. 

Energy-related phenomena in graphene span a wide range of energies, from optical down to THz. The high optical activity of 2D materials, which can absorb an order of magnitude more sunlight than Si layers of similar thickness,\cite{bernadi2013} is of interest for optoelectronics research.\cite{bonaccorso} 
Additionally,  the 2D structure 
renders electronic states fully exposed, allowing carriers and also heat to be extracted via a vertical transfer process (e.g. in a sandwich-type structure \cite{britnell1}). 
Another unique aspect of graphene is the ease with which electrons  can be pushed out of thermal equilibrium with the lattice. In such a hot-carrier regime, system states with elevated
 electronic temperatures different from that of the lattice can be fairly long-lived, resulting in electron energy transport decoupled from that of 
the lattice. 
The strong thermoelectric response of graphene \cite{zuev,wei} generates  strong coupling between energy modes and charge modes, giving rise to a range of 
 novel transport and optoelectronic phenomena.\cite{gabor,song,klaas,edrag,sun,folk14,junnanotech,herring} 
Strikingly, hot-carrier effects in graphene exist in a wide range of technologically relevant temperatures including room temperature.  Combined with its fast electrical response, this makes graphene an attractive material for high speed and gate tunable manipulation of energy flows on the nanoscale.

The hot carrier regime in graphene mainly stems from anomalously slow electron-lattice relaxation.\cite{wong,macdonald} 
Strong carbon-carbon bonds, which give graphene's lattice its rigidity, also result in a high optical phonon frequency, $\omega_0 = 200 \, {\rm meV}$. The large value of $\omega_0$ suppresses the optical phonon contribution to electron-lattice relaxation below a few hundred kelvin. At the same time, the generally weak scattering between electrons and long-wavelength acoustic phonons is further constrained by the large mismatch in Fermi velocity and sound velocity, $v/s \approx 100$. As a result, once the electrons are heated up they stay out of thermal equilibrium with the lattice over long times, proliferating over extended spatial lengthscales.\cite{macdonald,wong} 
Hot carriers have been studied in a variety of other systems including semiconductors like GaAs,\cite{shah,leadley,lutz} and metals.\cite{roukes} However in these other materials, hot carriers only exist at very low temperatures or under intense pumping. In contrast, hot carriers in graphene can exist even at room temperature and under weak driving.\cite{gabor}

The mechanisms responsible for the generation and cooling of hot carriers, which we discuss below, are characterized by vastly different time scales. For the carrier-carrier scattering processes occurring in the 
cascade triggered by photoexcitation, the times can be as fast as tens of femtoseconds.\cite{cavillieri,johannsen13} These fast times determine the branching ratio for the electron and phonon scattering pathways, controlling the energy part
captured by the electronic system when energy is pumped into graphene.\cite{klaas} The resulting hot-carrier state is manifested through an elevated electronic temperature.\cite{gabor,graham,grahamnano,sun,klaas,cavillieri,johannsen13} Once a hot carrier distribution is established, slower processes (up to hundreds of picoseconds), such as phonon emission, determine how long the carriers can stay hot. Both processes, fast scattering and slow cooling, contribute to the performance of optoelectronic devices.\cite{gabor,klaas,graham} 

Graphene's large interaction parameter, $\alpha \approx 2.2$, its 
gapless spectrum and tunable carrier density make it an ideal venue to probe the role of Auger type processes in the energy relaxation of high energy photoexcited carriers cascading down to lower energies. 
The large difference in the time scales for hot carrier generation and cooling allows us to treat these processes separately. Notably, as discussed below, graphene affords convenient knobs for controling both generation and cooling rates. These knobs provide means to manipulate the energy flows, underpining current attempts to exploit graphene as a future energy material.\cite{gabor,klaas}

A topically relevant context for studying energy flows in graphene is photoexcitation dynamics,\cite{kampfrath05,winzer,winzer12,winnerl11,george08,lui10, rana07,butscher07, vasko08, kim11,breusing09,klaas,cavillieri,johannsen13} which is sensitive to both the fast scattering processes of hot carrier generation and the slow electron-lattice cooling processes. 
Photoexcitation has been the main technique for probing hot carriers in graphene \cite{sun,gabor,graham,klaas,cavillieri,johannsen13} 
yielding high electronic temperatures under intense irradiation.\cite{lui10} 

In this article, we pedagogically lay out how energy flows through graphene electrons. To this end, we utilize the lens of hot carriers as a transparent way of addressing electronic energy flow in graphene. As we will see, this allows for a unified and multi-timescale treatment of photoexcitation that spans three orders of magnitude. 
In doing so, we survey relevant concepts introduced in the field, as well as supplement gaps with new results.
Many of the processes described herein work in synergy with each other and make graphene a favorable candidate for new optoelectronic devices. Indeed, utilizing some of the concepts reviewed in this article for efficient optoelectronics, including energy harvesting devices and photodetectors,\cite{herring,junnanotech,gabor} has become an emerging field. Here, we do not describe these engineering efforts in detail but instead focus on the physics of the fundamental processes.

The article is organized in two parts, discussing energy relaxation pathways at high and low energy (Sec.\,\ref{sec:pecascade} and 
Sec.\,\ref{sec:cooling}, respectively). As illustrated in Fig.\,\ref{pe}, this delineation also appropriately characterizes the time scales involved in the photoexcited carriers relaxation kinetics. 
 
In Sec.\,\ref{sec:pecascade}, we consider energy relaxation of (photoexcited) carriers at high energy. Focusing on the case of {\it doped} graphene, we detail the competition between carrier-carrier scattering (within a single band) and optical phonon emission. As we argue, the relatively high carrier density of doped graphene and large phase space available for carrier-carrier scattering allow it to dominate over optical phonon emission in the relaxation of high energy carriers. This manifests in a thermalized electron state that is characterized by an electronic temperature which is elevated above the lattice temperature. In this hot carrier regime, 
%
the carrier densities within each band remain unchanged; energy from the high energy electrons is captured by the ambient carriers as electronic heat. 

In Sec.\,\ref{sec:cooling}, we deal with thermal equilibration of the residual energy captured by ambient carriers. The relevant processes - electron-lattice cooling - occur at low energies, relaxing electrons close to the Fermi surface; we discuss these processes near equilibrium, giving estimates of the cooling power for different pathways.
The generally low efficiency of these pathways leads to an interesting competition problem. 
As we will see, while first-order processes such as single optical and acoustic phonon emission are ineffective, 
there are other processes -- {\it supercollisions} -- that can relieve the cooling bottleneck. We discuss disorder-assisted supercollisions, \cite{song12} recently seen by a number of experimental groups,\cite{graham,betz2013,qiong,grahamnano} as well as other forms of supercollisions and their physical manifestations. These include {\it few-body} scattering off ripples, flexural phonons, and pairs of counterpropagating phonons. 


\section{Hot Carrier Generation and the Photoexcitation Cascade}
\label{sec:pecascade}

We begin by examining the photoexcitation cascade in graphene with emphasis on how energy flows and gets partitioned between different degrees of freedom. As we will see, the generation of hot carriers is an important process contributing to 
energy relaxation after initial photoexcitation. Here we will concentrate on the experimentally relevant case of doped graphene.  

\begin{figure}
\includegraphics[scale=0.26]{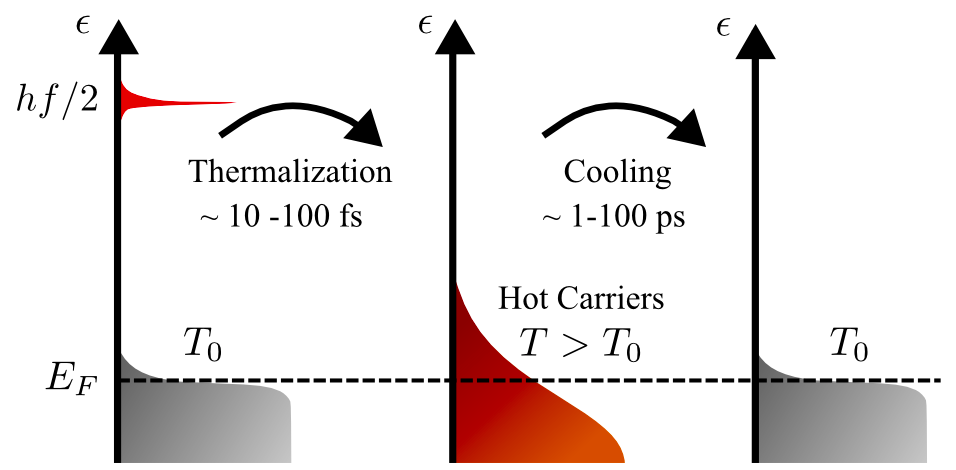}
\caption{Main stages of energy relaxation of photoexcited carriers. 
In the first stage, 
the carriers cascade from energy $\epsilon = hf/2$ down to low energies, losing energy via Auger processes and phonon emission. These processes lead to fast thermalization over timescales on the order tens to hundreds of femtoseconds, producing a relatively long-lived hot carrier distribution (middle panel).
In the second stage, electron-lattice cooling mediated by acoustic phonons takes place over longer time scales (several to a hundred picoseconds) relaxing the hot carrier distribution back to equilibrium ($T=T_0$, right panel).}
\label{pe}
\end{figure}

Photoexcitation in doped graphene proceeds as depicted in Fig.\,\ref{pe}: photons with energy $hf > 2\mu$
create high-energy electron-hole pairs
which form an outlying distribution of carriers high above the Fermi level (small red peak in Fig.\,\ref{pe} left panel); $\mu$ is the chemical potential. \cite{cavillieri,johannsen13, george08} 
A similar peak is formed by holes in the negative-energy band. 
The high-energy carriers then relax, losing energy to phonons or scattering with ambient carriers.\cite{kampfrath05,winzer,winzer12,winnerl11,george08,lui10, rana07, butscher07, kim11, breusing09, klaas, pollini,cavillieri,johannsen13} 
The amount of energy captured by the electronic system in this fast thermalization process depends on the competition between the rates for these pathways, which we will discuss below. 
After thermalization, a hot carrier distribution is formed with an elevated electronic temperature (Fig.\,\ref{pe} middle panel). This hot carrier distribution subsequently cools down by the emission of acoustic phonons to the lattice.\cite{macdonald,wong,song12,graham,betz2013,grahamnano} In this section, we discuss the thermalization process (hot carrier generation). In the next section we discuss how the hot carrier distribution cools with the lattice.

A central question in the thermalization cascade is how the energy of photoexcited carriers is partitioned between the electron and lattice degrees of freedom. 
There are two primary types of energy relaxation 
pathways: (i) carrier-carrier scattering\cite{rana07,ie,winzer,winzer12,klaas} and (ii) optical phonon emission.\cite{kampfrath05,butscher07,kim11} Only the processes of type (i) help to produce hot carriers, since the energy lost to the lattice degrees of freedom has very little impact on electron temperature (the lattice specific heat can be $10^3-10^4$  times larger than electron contribution, hence the energy transferred directly to electron is much more effective in producing hot carriers). The rates for these pathways analyzed in recent literature were found to be quite fast  (tens to hundreds of fs), lying in the same ballpark for processes of type (i) and (ii).\cite{ie} The competition of these pathways determines the amount of energy that the ambient distribution captures from the initially photoexcited high-energy carriers. 

Type (i) processes, which are key for various graphene-based energy harvesting proposals, have recently received a lot of attention.\cite{rana07,ie,winzer,winzer12,klaas} Microscopically, they can be understood as Auger type processes in which carrier-carrier scattering between photoexcited high energy carriers scatter with low energy ambient carriers, see Fig.\,2.  
The technological promise of graphene energy harvesting largely relies on the high effectiveness of these processes in passing energy among the high- and low-energy carriers. 
To avoid confusion, we differentiate the Auger processes into two distinct classes namely: a) intra-band carrier-carrier scattering (also called Impact Excitation, Auger Heating, see Fig. \ref{auger}a) and b) inter-band carrier-carrier scattering (also called Impact Ionization, Carrier Multiplication, see Fig.\,\ref{auger}b). Importantly, while intra-band carrier-carrier scattering (Fig. \ref{auger}a) conserves the number of carriers in each band, inter-band carrier-carrier scattering (Fig. \ref{auger}b) does not. 
However, as discussed below, inter-band carrier-carrier scattering is blocked by kinematic constraints arising due to energy and momentum conservation, rendering intra-band scattering the dominant mechanism for hot carrier production (Fig.\,\ref{pe} middle panel). In addition to scattering processes a) and b), there also exist electron-hole recombination processes arising from carrier-carrier scattering (Auger recombination). These processes were analyzed in Refs. \onlinecite{rana07,rana09,winzer,winzer12}; here we will concentrate on a) and b). 

We begin with the Hamiltonian for graphene $N=4$ species of massless Dirac carriers,
\bea\label{eq:H}
&&
\mathcal{H}=  \sum_{\vec{k},i}  \psi^\dag_{\vec k,i}  H_0 \psi_{\vec k,i}  + \mathcal{H}_{\rm el-el} +  \mathcal{H}_{\rm el-ph} +  \mathcal{H}_{\rm ph}
, 
\\\nonumber 
&&
\mathcal{H}_{\rm el-el} = \frac12\sum_{\vec q, \vec k,\vec{k'},i,j}V(\vec q) \psi^\dag_{\vec k + \vec q, i} \psi^\dag_{\vec{k'} - \vec q,j} \psi_{\vec {k'},j} \psi_{\vec k, i} 
.
\eea
Here $\psi_{\vec k,i}$ $\psi_{\vec k,i}^\dagger$ describe two-component (pseudo)spin states, $i,j=1...N$ label valley and spin degrees of freedom. The term $H_0 = v\sigma \cdot \vec k$ describes graphene's Dirac spectrum, $V(\vec q)=2\pi e^2/|\vec q|\kappa$ is the Coulomb interaction, and $\kappa$ is the bare dielectric constant. The terms $\mathcal{H}_{\rm ph}$ and $\mathcal{H}_{\rm el-ph}$ describing phonons in graphene and the electron phonon interaction will be specified below. 
In the above Hamiltonian, Eq.(\ref{eq:H}), we ignored several effects some of which are small and some of which will be introduced later as needed. The small effects are intervalley scattering terms in the carrier-carrier interaction and Umklapp scattering. The former is small because it originates from short range Hubbard-type interactions, which are smaller than the long range $V(\vec q)$ interaction. Umklapp scattering in graphene can only occur due to three-particle collisions. This is so because, due to graphene lattice symmetry, vectors $\vec K$, $\vec K'$ and $\vec K-\vec K'$ are 3 times smaller than the reciprocal lattice vectors.
In the single-particle Hamiltonian $H_0$ we ignored effects such as coupling to disorder and trigonal warping of the linear Dirac spectrum. Each of these effects will be discussed in due time.

\subsection{Carrier-Carrier Scattering: Kinematic Constraints}

\begin{figure}
\includegraphics[scale=0.2]{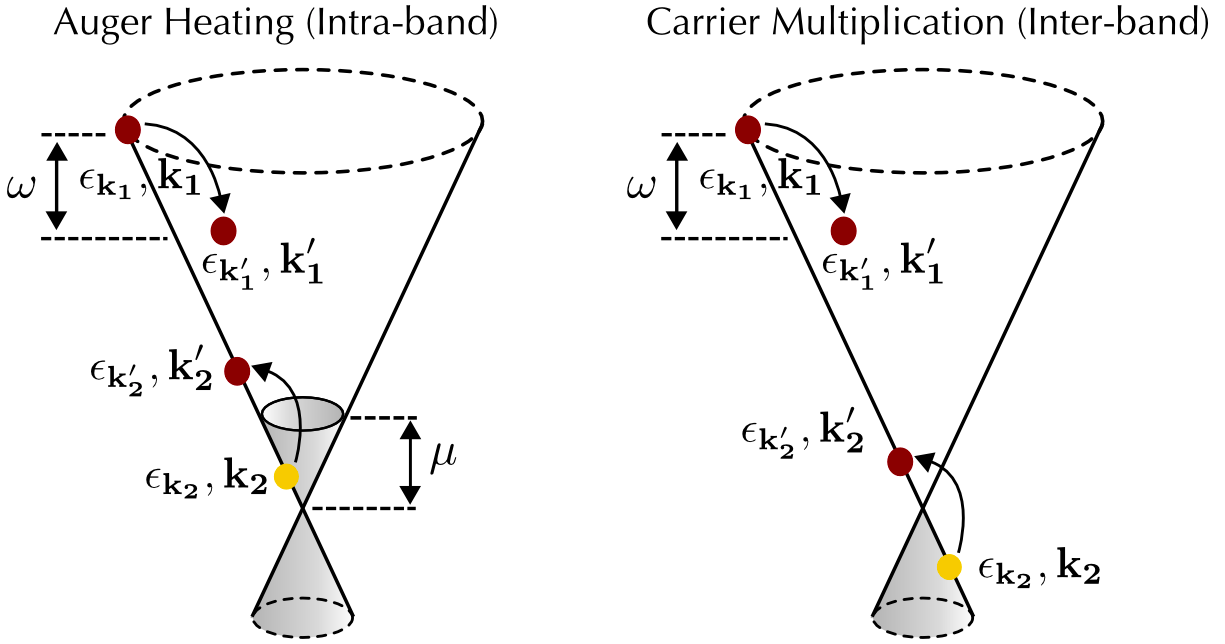}
\caption{Types of carrier-carrier scattering in graphene: a) Intra-band processes,
and b) Inter-band processes. Processes a) may change electron temperature (hot carrier generation) but cannot change the number of carriers in each band. Carrier multiplication may only occur due to processes b) which change the carrier number in a band. The transition rate, described by Eq.(\ref{eq:W}), obeys kinematic constraints due to energy and momentum conservation. 
Because of these constraints, processes b)
can only occur when transitions are collinear, $\omega = v|\vec q|$, resulting in a severely constrained phase space effectively blocking b) for two-body collisions (see text).
}
\label{auger}
\end{figure}

We first consider Auger-type processes depicted in Fig.\,\ref{auger}. Both these processes involves a photo-excited carrier with high energy and momentum, $\epsilon_\vec{k_1}\gg \mu$, $|\vec k_1|\gg k_{\rm F}$,  which is scattered to a lower energy state of momentum  $\vec k_1'$ with recoil momentum $\vec q = \vec{k_1} - \vec k_1'$ passed to an  electron in the Fermi sea.  Here $k_{\rm F}$ is the Fermi wavevector. The latter particle-hole pair excitation 
process is 
depicted by a transition from $\vec{k_2}$ to $\vec k_2'$ in Fig.\,\ref{auger}. The transition rate for this process, evaluated by the 
Golden Rule  approach, takes the form \cite{ie,tse08}
\bea\label{eq:W}
W_{\vec{k_1'},\vec{k_1}}  &=& \frac{2\pi N}{\hbar}\sum_{\vec q, \vec{k_2}, \vec{k_2'}} 
 f_\vec{k_2}( 1- f_{\vec{k_2'}}) F_{\vec{k_2}, \vec{k_2'}'}| \tilde V_\vec{q} |^2
\\\nonumber
&&\times \delta_{\vec{k_1'}, \vec{k_1} + \vec q} \delta_{\vec{k_2'}, \vec{k_2} - \vec q}  \delta (\epsilon_{\vec{k_1'}}-\epsilon_{\vec{k_1}}  + \epsilon_{\vec{k_2'}}-\epsilon_{\vec{k_2}})
.
 \eea
Here $f_{\vec k}$ is a Fermi function, and $ F_{\vec k, \vec k'}=|\la \vec k's'|\vec k s\ra|^2$ is the coherence factor ($s,s'=\pm$ label states in the electron and hole Dirac cones).  The effective Coulomb interaction $\tilde{V}$ which mediates scattering between the photo-excited carrier and the carriers in the Fermi sea
is taken in the form
\be
\tilde V_\vec{q} = \frac{V_{\vec q}^0}{\varepsilon(\omega,\vec q)}
,\quad
\varepsilon(\omega,\vec q)=1- V_\vec{q}^0 \Pi (\vec q, \omega)
.
\ee
Here $V_{\vec q}^0 = 2\pi e^2/|\vec q|\kappa$, and the ``permittivity'' $\varepsilon(\omega,\vec q)$ accounts for dynamical screening. This random-phase approximation (RPA) model uses 
the polarization operator
$
\Pi (\vec q, \omega) =N\sum_{\vec k, s, s'} F_{\vec{k}, \vec{k} + \vec{q}; ss' } \frac{ f(\epsilon_{\vec{k}, s}) - f(\epsilon_{{\vec{k}+\vec q}, s'})}{ \omega + \epsilon_{\vec{k}, s}- \epsilon_{\vec{k} + \vec{q}, s'}  + i0}
,
$
with the band indices $\{s, s'\} = \pm $. This includes both intra- ($s=s'$) and inter- ($s\neq s'$) band contributions.\cite{dassarma} 

Importantly, the electronic transitions in a massless Dirac band governed by the Hamiltonian (\ref{eq:H}) are subject to {\it kinematic constraints}.\cite{aleiner,rana07} These constraints arise due to the combined effect of linear dispersion in two Dirac cones, $E_{\pm}(\vec p)=\pm v|\vec p|$, and the momentum conserving character of carrier scattering. 
Here we analyze the simplest case of a two-body collision.
Each of the two particles participating in a collision can make transitions between states in the upper and lower Dirac cones which we denote by $+$ and $-$. Two kinds of transitions can be distinguished: intra-band transitions ($+\to +$ or $-\to -$), see eg. Fig.\,\ref{auger}a yellow to red circles, and inter-band transitions ($+\to -$ or $-\to +$), see eg. Fig.\,\ref{auger}b yellow to red circles. Since momentum change in any transition satisfies $||\vec k|-|\vec k'||<|\vec k-\vec k'|<|\vec k|+|\vec k'|$, intra-band transitions can only occur when the energy and momentum change are related by $|\Delta\epsilon|\le v|\Delta k|$, whereas inter-band transitions are possible only when $|\Delta\epsilon|\ge v|\Delta k|$. Here $\vec k$ and $\vec k'$ denote initial and final momentum for a single transition; $\vec k \to \vec k'$ can refer to either of the transitions $\vec k_1\to  \vec k_1'$ or $\vec k_2 \to \vec k_2'$ as shown in Fig.\,\ref{auger}.

For Eq.(\ref{eq:W}) to give a non-vanishing result, the transitions $\vec k_1\to \vec k_1'$, $\vec k_2\to \vec k_2'$ must occur in like pairs, i.e. both transitions are intra-band or both are inter-band. Since the transition $\vec k_1\to \vec k_1'$  is restricted to be within a single band, the transition $\vec k_2\to \vec k_2'$ must also be intra-band. While inter-band carrier-carrier scattering in Fig.\,\ref{auger}b can technically occur when the energy and momentum exchanged are collinear $\omega = v|\vec q|$, the vanishing phase space for these transitions effectively block inter-band carrier-carrier scattering.
As a result, intra-band carrier-carrier scattering in Fig.\,\ref{auger}a are expected to play the dominant role in Auger-type relaxation in doped graphene.\cite{ie,klaas,cavillieri,johannsen13} 

Kinematical blocking of inter-band processes can in principle be relieved by three-body (or, higher-order) collisions. Indeed, since the constraints arise from simultaneous momentum and energy conservation, relaxation of momentum conservation for example via coupling to disorder or high order processes in which multiple pairs are created at the same time provide a viable route to unblocking inter-band carrier-carrier scattering. Such processes may become important at high excitation power, however we expect the effect of such processes to be weak in the low excitation power regime discussed below. 

Finding pathways in which interband processes are allowed has been the subject of recent research; several authors have investigated inter-band carrier-carrier scattering both theoretically and experimentally.\cite{winzer,winzer12,pollini,cavillieri,johannsen13,tomadin} Early theoretical work in Ref. \onlinecite{winzer,winzer12} that simulated the early thermalization dynamics of photoexcited carriers through a density matrix formalism and Bloch equations suggested that inter-band processes were possible. In particular, Ref. \onlinecite{winzer,winzer12} predicted that Carrier Multiplication events dominated over Auger Recombination particularly under high excitation power; this in part spurred much of the current interest in Auger processes in graphene.\cite{bonaccorso} Later, other theories suggested that inter-band carrier-carrier scattering events are allowed when electron lifetime effects,\cite{tomadin} or when trigonal warping was taken into account.\cite{baskotrigonal} On the experimental end, Angle Resolved Photoemission Spectroscopy (ARPES) \cite{cavillieri,johannsen13} and optical pump THz probe \cite{pollini} experiments have been used to search for 
carrier multiplication events. 
However, differences in techniques and results have left the current status of interaction mediated relaxation in undoped graphene hotly contested.\cite{pollini, cavillieri,johannsen13} Since the intraband process in Fig.\,\ref{auger}a vanishes in undoped graphene, a fuller understanding of the high excitation power regime, how higher order collisions affect interband scattering, and how these Auger processes compete with phonon emission is needed for a complete picture of the photoexcitation cascade in undoped graphene. 

Below, we focus on the technologically relevant case of doped graphene. In this case, intra-band carrier scattering is both kinematically allowed and can provide fast relaxation of photoexcited carriers.\cite{ie} Importantly, the intra-band scattering processes allow the ambient carrier distribution to capture the energy of photoexcited carriers as electronic heat, elevating the electronic temperature and giving rise to the hot carrier regime (middle panel, Fig.\,\ref{pe}). In doped graphene, energy captured via hot carriers becomes the most relevant quantity for evaluating the efficiency of graphene optoelectronics since strong thermoelectricity in graphene \cite{zuev,wei} allow hot carriers to drive optoelectronic circuits, and dominates its photocurrent response.\cite{gabor}

\subsection{Intra-band Carrier-Carrier Scattering}
\label{sec:ie}

The energy relaxation rate of a photo-excited carrier via intra-band carrier-carrier scattering is \cite{ie}
\be
\mathcal{J}_{\rm el} =\sum_{\vec{k_1'}} (\epsilon_{\vec k_1'} - \epsilon_{\vec k_1}) W_{\vec k_1', \vec k_1} ( 1- f_{\vec k_1'})  F_{\vec{k_1}, \vec k_1' } 
\ee
As shown in Fig.\,\ref{auger}a, the typical energy of an excited pair is much smaller than the photo-excitation energy  $\epsilon_{\vec k_1}$. As a result, it is convenient to factorize the transition rate through the spectrum of secondary pair excitations using the secondary pair susceptibility, $\chi''(\vec q, \omega) = N\sum_{\vec k} F_{\vec{k}, \vec{k} + \vec{q} } (  f_{\vec{k}} - f_{\vec{k}+\vec q}) \delta(\epsilon_{\vec{k+q}}-\epsilon_{\vec{k}} - \omega) =  - \frac1{\pi}{\rm Im} \Pi_+(\vec q,\omega)$; for the intra-band process discussed below $\Pi_+(\vec q, \omega)$ refers to polarization from intra-band scattering only. Following this standard procedure, we obtain
\bea\label{eq:Gamma_total}
\mathcal{J}_{\rm el} (\epsilon_{\vec k_1}) &&=\int_{-\infty}^{\infty}\!\!\! d\omega  \omega P(\omega)
,
\\ 
P (\omega) &&=
A \sum_{\vec q} | \tilde V_{\vec{q}} |^2 F_{\vec{k_1}, \vec k_1' }\chi''(\vec q, \omega) 
\delta(\epsilon_{\vec k_1'}-\epsilon_{\vec k_1} +\omega)
,
\label{eq:p}
\eea
where $A=\frac{2\pi}{\hbar}[N(\omega)+1)][ 1 - f(\epsilon_{\vec k} - \omega)]$ and $\vec k_1'=\vec k_1-\vec q$, and $\omega$ denotes the amount of energy transferred in each scattering event, and $P(\omega)$ is the transition probability for scattering. 

Interestingly, $P(\omega)$ depends very weakly on the initial energy of the original photoexcited pair, since $\epsilon_\vec{k}$ only enters in the Fermi function in $A$. Further, numerically evaluating Eq.(\ref{eq:p}), Ref. \onlinecite{ie} found a non-monotonic $P(\omega)$ peaking close to $ \omega \approx \mu$, and decaying rapidly for $\omega \gg \mu$.\cite{ie} As a result, for energies $\epsilon_\vec{k_1} \gg \mu$, the energy relaxation of a high energy carrier dominated by carrier-carrier scattering proceeds in steps of $\mu$. 

Large values of $\mathcal{J}_{\rm el}$, as large as several ${\rm eV}/ {\rm ps}$, have been estimated for typical dopings in graphene.\cite{ie} The efficiency of intraband carrier-carrier scattering noted above can be linked to the large values of $\mu$ in graphene. Indeed, the relation between efficiency and $\mu$ can be clarified by simple dimensional analysis. The weak dependence of $P(\omega)$ on initial carrier energy (see above) means that it depends on $\omega$ essentially via the dimensionless parameter $x=\omega/\mu$. As a result, $\mathcal{J}_{\rm el}$ can be described by
\be
 \mathcal{J}_{\rm el}(\epsilon) = \frac{\mu^2}{\hbar} \int_0^{\epsilon/\mu} \, x\tilde{P}(x) dx,
\label{eq:Jnumerical}
\ee
where $\tilde P(x) = \hbar P(\Delta \epsilon)$ is dimensionless. Efficient $\mathcal{J}_{\rm el}$ arises from the fast $\Gamma \sim \mu/(2\pi\hbar) \approx 20 \, {\rm ps}^{-1}$  (for typical doping of $\mu  =  0.1 \, {\rm eV}$) allowed from unitarity; the large density of carriers available within the band in doped graphene provides a large phase space for fast intraband carrier-carrier scattering. 

\subsection{Optical Phonon Emission}
\label{sec:opticalphonons}

An alternative channel for energy relaxation of photo-excited carriers occurs through the emission of optical phonons and gives an energy relaxation rate of  $\mathcal{J}_{\rm ph}$. The transition rate of this process \cite{tse08} can be described by Fermi's golden rule
\bea
W_{\vec{k}',\vec{k}}^{\rm el-ph}  &=& \frac{2\pi N}{\hbar}\sum_{\vec q} |M(\vec{k}',\vec{k})|^2  \\ \nonumber
&& \delta \big(\Delta\epsilon_{\vec{k}', \vec{k}}  + \omega_\vec{q}  \big) \delta_{\vec{k}', \vec{k} + \vec q} (N({\omega_\vec{q}}) + 1),
\eea
where $\Delta\epsilon_{\vec{k}', \vec{k}} = \epsilon_{\vec k'} - \epsilon_{\vec k}$, $\omega_\vec{q}=\omega_0 = 200\, {\rm meV}$ is the optical phonon dispersion relation, and $N({\omega_{\vec q}})$ is a Bose function. Here $\vec k$ is the initial momentum of the photo-excited electron, $\vec k'$ is the momentum it gets scattered into, and $\vec q$ is the momentum of the optical phonon. The electron-phonon matrix element $M(\vec{k}',\vec{k})$ is  \cite{andoph,macdonald,wong}
\be
|M(\vec{k}',\vec{k})|^2 = g_0^2 F_{\vec{k}, \vec{k}' }, \quad g_0 = \frac{2\hbar^2v}{\sqrt{\rho \omega_0 a^4}},
\ee
where $ F_{\vec{k}, \vec{k}' }$ is the coherence factor for graphene, $g_0$ is the electron-optical phonon coupling constant\cite{note},  
$\rho$ is graphene's mass density, and $a$ is the distance between nearest neighbor carbon atoms. In the same fashion as above, the energy-loss rate of the photo-excited carrier at energy $\epsilon$ due to the emission of an optical phonon can be evaluated as 
$
\mathcal{J}_{\rm ph}(\epsilon) = \sum_{\vec{k'}} W_{\vec{k}', \vec{k}}^{\rm el-ph} (\epsilon_k' - \epsilon)\big[ 1- f(\epsilon_\vec{k'})\big]
$. This yields the rate \cite{ie}
\be
\label{eq:Jph}
\mathcal{J}_{\rm ph}(\epsilon) = \frac{\pi N}{\hbar} \omega_0 g_0^2 \big[ 1- f(\epsilon - \omega_0)\big]  (N(\omega_0)+1)\nu(\epsilon - \omega_0), 
\ee
where $\nu(\epsilon)=\epsilon/(2\pi v^2\hbar^2)$ is the electron density of states in graphene. $\mathcal{J}_{\rm ph}(\epsilon)$ varies linearly with the photo-excited carrier energy $\epsilon > \omega_0 + \mu$ and vanishes for $\epsilon<\omega_0+\mu $; here we have set $T=0$ for clarity (a good approximation for $k_{\rm B}T \ll \mu, \omega_0$). Because the electron-optical phonon coupling is a constant, this result is to be expected from the increased phase space to scatter into at higher photo-excited carrier energy. 

To get an order of magnitude estimate of the energy relaxation rate, we use values $(N(\omega_0)+1) \approx 1$ and $1- f(\epsilon - \omega_0) \approx \Theta(\epsilon - \omega_0- \mu)$. This gives
\be
\mathcal{J}_{\rm ph} (\epsilon) 
\approx \frac{\epsilon - \omega_0}{\tau_0} \Theta(\epsilon - \omega_0- \mu), \quad \tau_0 = \frac{2v^2 \hbar^3}{N \omega_0 g_0^2}
\label{eq:estimateph}
\ee
Using $\rho = 7.6 \times 10^{-11} \, {\rm kg} \, {\rm cm}^{-2}$,
we find $\tau_0 \approx 350 \, {\rm fs}$. 

\subsection{Auger Processes vs. Phonon Emission: Branching Ratio}

Comparing $\mathcal{J}_{\rm el}$ from intraband carrier-carrier scattering (in Sec.\,\ref{sec:ie}) with the energy relaxation rate arising from the emission of optical phonons, $\mathcal{J}_{\rm ph}$ (in Sec.\,\ref{sec:opticalphonons}), yields a branching ratio $\mathcal{J}_{\rm el}/\mathcal{J}_{\rm ph}> 1$ for typical dopings.\cite{ie} Indeed, large branching ratios $ \mathcal{J}_{\rm el}/\mathcal{J}_{\rm ph} \sim 4$ have been inferred experimentally.\cite{klaas} 
Furthermore, while $\mathcal{J}_{\rm ph}$ does not depend on carrier density, $\mathcal{J}$ does. As a result, gate voltage 
can be used to tune the branching ratio $\mathcal{J}_{\rm el}/\mathcal{J}_{\rm ph}$.\cite{ie} 

The amount of heat absorbed, $\Delta Q_{\rm el}$, by the electronic system from the cascade of a single photoexcited carrier at $\epsilon = hf/2$ is
\be
\Delta Q_{\rm el} = \int_0^{t_0} \mathcal{J}_{\rm el} dt = \int_{\mu}^{hf/2} \frac{d\epsilon}{1 + \mathcal{J}_{\rm ph}/\mathcal{J}_{\rm el}}.
\ee
For large enough $\mu \gtrsim 0.1$, $\mathcal{J}_{\rm el}/\mathcal{J}_{\rm ph} > 1$. As a result, a large fraction of the energy from the photoexcited carriers is absorbed as heat in the electronic system. This precipitates an increase in the electronic temperature, $T_{\rm el}$, and generates hot carriers.  

\subsection{Hot Carriers for Energy Harvesting}

The ability to drive optoelectronic circuits from an elevated $T_{\rm el}$, \cite{song,gabor} means that the energy captured by fast carrier-carrier scattering can be used to increase the efficiencies of optoelectronic devices. Indeed, one proposal in which efficiencies may be gained is using graphene's unique hot carrier generating characteristics for ``hot carrier solar cells''.\cite{nozik} In these types of solar cells, the extraction of hot carriers can be used to drive a nanoscale heat engine to reach efficiencies as high as $60 \, \%$, far larger than that imposed by the more traditional Shockley-Queisser limit for single-junction photovoltaic cells.\cite{sqlimit} 

There are two material requirements for such solar cells: i) efficient thermalization of high energy photoexcited electrons by ambient carriers so that photon energy can be captured as electronic heat, and ii) slow electron-lattice cooling so that the heat captured in (i) is not quickly lost to the lattice.\cite{nozik} Indeed, the inefficient thermalization in conventional solar cells (such as those made out of Silicon) is one of the largest contributors to the low efficiencies of photovoltaic cells limited by the Shockley-Queisser limit. \cite{sqlimit}

The combination of fast intraband Auger processes described in this section, slow electron-lattice cooling \cite{wong,macdonald} that can be achieved in high quality graphene, and graphene's two-dimensional nature which enable vertical extraction of hot carriers make it an interesting candidate material for a new paradigm of solar cells based on hot carriers.\cite{nozik} Vertical extraction is a front runner among the currently discussed proposals for extracting hot carriers to drive optoelectronic circuits;\cite{gabor,britnell1,junnanotech,herring} utilizing graphene in energy harvesting is a topic of active research. 
 
\subsection{Measuring Hot Carrier Temperature}

Measuring the electronic temperature is an ideal way of experimentally tracking energy flows in graphene. Since fast carrier-carrier scattering around the Fermi surface (tens of femtoseconds) allows for an electronic temperature to be established quickly \cite{cavillieri,johannsen13,george08} (see also above discussion on intra-band carrier-carrier scattering), the electronic temperature can be used to characterize both the short timescale thermalization and longer timescale cooling. Indeed, tracking the temperature dynamics of hot carriers \cite{graham} provides a sensitive probe of the cooling mechanisms that will be discussed in the next section.  
There are a variety of ways of measuring electronic temperature in graphene, including photocurrent,\cite{graham,sun} noise-thermometry,\cite{betz2013} transient absorption,\cite{grahamnano} angle resolved photoemission,\cite{cavillieri} and THz conductivity.\cite{choi09,hwangnonlinear,rana,klaas,sufei,alex,frenzel13,heinz13} Recently, the last two have been used in pump-probe type experiments \cite{cavillieri,johannsen13,klaas} to probe the initial thermalization dynamics of the photoexcitation cascade (Fig.\,\ref{pe}, left panel). Photoconductivity in pump-probe type experiments have recently received wide attention because of their ability to probe a wide variety of dynamical processes that can occur on short time scales.\cite{ulbricht} Here, we will discuss THz conductivity and how effects from the dynamics of the electronic temperature alone lead to changes in the measured THz conductivity. 

As we will see, the optical conductivity depends directly on the energy-dependent
carrier distribution, $n_\vec{k}$, and gives a different value according to how
far away from equilibrium the carrier distribution is. Analysis of a kinetic equation using the relaxation time approximation produces the conductivity 
\be \label{eq:conductivity} \sigma (\omega) =-\sum_{\vec
k}\frac{\tau_{\rm tr}(\epsilon_{\vec k})}{1+ [\omega\tau_{\rm
tr}(\epsilon_{\vec k})]^2}e^2\vec v_{\vec k} \cdot \nabla_{\vec k} n_{\vec k}
\ee
where ${\tau_{\rm tr}(\vec k)}$ is the transport scattering time, $\vec v_{\vec k} = \partial \epsilon_\vec{k}/\partial \vec k$ is the group velocity, and $n_{\vec k} = (1 + e^{\beta (\epsilon_{\vec k} - \mu)})^{-1}$ depends on the electronic temperature through $\beta = 1/ k_{\rm B} T_{\rm el}$. Here we have shown only the real part of the conductivity readily obtained in THz photoconductivity experiments;\cite{choi09,hwangnonlinear,rana,klaas,sufei,alex,frenzel13,heinz13} the imaginary part can also be similarly obtained. In pump-probe type experiments, the pump-induced change in conductivity (also called photoconductivity) provides information about how far the carrier distribution is pushed out of equilibrium. As such, we will be most interested in the temperature dependent changes in conductivity $\Delta \sigma_{\omega}  = \sigma_{\omega}(T_{\rm el, 1}) -  \sigma_{\omega}(T_{\rm el, 0})$. Here $T_{\rm el,0}$ denotes the initial electronic temperature and $T_{\rm el, 1}$ denotes the electronic temperature after pump. 

Two regimes naturally arise in graphene: (i) the degenerate limit, $\mu \gg k_{\rm B}T_{\rm el}$, in doped graphene, and (ii) the non-degenerate limit, $\mu < k_{\rm B}T_{\rm el}$, in graphene close to charge neutrality. Since carrier density in graphene can be tuned by gate, these two regimes can be easily accessed. 

In the degenerate limit, $\mu \gg k_{\rm B}T_{\rm el}$, a Sommerfeld expansion can be employed in the analysis of Eq.(\ref{eq:conductivity}). This yields an estimate for $\Delta \sigma_{\omega}$ as \cite{klaas}
\bea
\label{eq:degenerate}
\Delta \sigma_{\omega} &&\approx  \Big( k_{\rm B}^2 T_{\rm el, 1}^2 -
k_{\rm B}^2 T_{\rm el, 0}^2\Big) \frac{\pi^2}{6}
N \nu(E_F)\frac{\partial^2F(\epsilon)}{\partial
\epsilon^2}\Bigg|_{\epsilon=E_F} \nonumber \\
F(\epsilon) &&= e^2 v^2 \frac{\tau_{\rm tr}(\epsilon)}{1+\omega^2[\tau_{\rm tr}(\epsilon)]^2} 
\eea
where $E_F$ is the Fermi energy, and $N=4$ is the valley/spin degeneracy in graphene. Here the carrier density has been fixed constant by accounting for changes in chemical potential as a function of temperature $\mu \approx E_F - \frac{\pi^2}{6} k_{\rm B}^2 T^2 / E_F$; temperature dependence of the transport time, $\tau_{\rm tr}(\epsilon)$, has also been neglected and can be included in a more elaborate analysis. Both of these assumptions are valid when $\mu \gg k_{\rm B}T_{\rm el}$. 

For graphene, the transport time can be modeled as $\tau_{\rm tr} = a \epsilon$;\cite{nomura, ando,dassarmareview} here $a$ is a sample dependent constant. As a result, $\frac{ \partial^2 F(\epsilon)}{\partial \epsilon^2} < 0$ for
small frequencies $\omega \sim {\rm THz}$ giving a {\it negative} $\Delta \sigma_\omega < 0$ when electronic system gets hotter $T_{\rm el, 1} > T_{\rm el, 0}$. Here, an estimate of sample dependent $a$ can be extracted from the dc conductivity via the einstein relation, $\sigma_{\rm dc} = e^2 v^2 \nu(\mu) \tau(\mu) / 2$.
We note parenthetically, that since the heat capacity in the degenerate limit goes as $C_{\rm el} \propto T$, the absolute change in optical conductivity $|\Delta \sigma_{\omega} | \propto \Delta Q_{\rm el}$ making the optical conductivity a good probe of the amount of heat injected into the electronic system. In the context of probing the photoexcitation cascade, $|\Delta \sigma_{\omega} |$ is sensitive to the amount of heat captured by the ambient carriers.\cite{klaas} Pump induced negative photoconductivity in doped graphene has recently been observed experimentally.\cite{klaas,frenzel13,heinz13,hwangnonlinear} 

In contrast, $\Delta \sigma_{\omega} > 0$ in the non-degenerate limit ($\mu < k_{\rm B}T_{\rm el}$). To illustrate this, we use $\tau_{\rm tr} = a \epsilon$ as above and fix $\mu=0$. For $\omega \tau_{\rm tr}(\epsilon = k_{\rm B}T)  \ll 1$, we estimate
\be
\label{eq:non-degenerate}
\Delta \sigma_{\omega} \approx \frac{Ne^2 a \pi}{12 \hbar^2} k_{\rm B}^2 \Big (T_{\rm el, 1}^2- T_{\rm el, 0}^2 \Big )  
\ee
which is {\it positive} as the temperature of the electronic system gets hotter $T_{\rm el, 1} > T_{\rm el, 0}$. Interestingly, $\Delta \sigma$ in this case is no longer directly proportional to the amount of heat pumped in the the system (since $C_{\rm el} \propto T^2$ in the non-degenerate limit). In recent experiments, the different signs of the two regimes have been observed in dynamical measurements involving an optical pump THz probe of a single gate tunable graphene sample.\cite{sufei,alex} These have been described in terms of metallic behavior (in the degenerate limit) and semiconducting-like behavior (non-degenerate limit) respectively.\cite{sufei,alex}  A related gate tunable photoconductivity sign change was also observed in steady state photoconductivity measured in biased graphene.\cite{avourisphotoconductivity}

While we only describe the effect of an increase in electron temperature on $\Delta \sigma$, a number of other scattering channels can also affect $\Delta \sigma$.\cite{perebeinos,klaas,sufei,alex} However, many of the qualitative features, such as photoconductivity sign, remain unchanged. For example, using an energy independent $\tau_{\rm tr}$, Ref. \onlinecite{sufei,alex} also found a positive $\Delta \sigma$ in the non-degenerate limit; positive $\Delta \sigma$ in this limit has been attributed to a change in the Drude weight.\cite{sufei,alex} Additionally, optical phonons in both graphene \cite{perebeinos,klaas} and the substrate \cite{perebeinos} emitted in the photoexcitation cascade can also scatter with ambient carriers to change the optical conductivity measured. 
%

\section{Hot Carrier Cooling and Supercollisions}
\label{sec:cooling}

The mechanisms responsible for cooling hot carriers in graphene control how long the electronic system stays out of thermal equilibrium with the lattice. The time scales over which the electron system is hotter than the lattice determines the time scales (or length scales) over which hot carrier effects can be observed.\cite{song} 
For instance, large and spatially extended photocurrent response in graphene p-n junctions has been attributed to long hot carrier cooling lengths.\cite{gabor} As a result, hot carrier cooling mechanisms play a critical role in determining the magnitude and qualitative behavior of graphene response. Below we detail what controls cooling in graphene. 

Hot carrier cooling proceeds via the emission of phonons through the electron-phonon interaction $\mathcal{H}_{\rm el-ph}$ in Eq.(\ref{eq:H}). 
The electron-phonon interaction in graphene arises from the usual deformation potential coupling, 
\be
\label{eq:H_elph}
\mathcal{H}_{\rm el-ph} = \sum_{\vec q} g \sqrt{\omega_{\vec q}}\big( b_\vec{q} + b^\dag_{-\vec q}\big) n_{\vec q},\quad
g = D/\sqrt{2\rho s^2}
,
\ee
where $D$ is the deformation potential, $\rho$ is the mass density of the graphene sheet, and $s$ is the speed of sound in graphene. In this section, we use $D = 20\, {\rm eV}$ throughout.\cite{macdonald,wong} As discussed earlier, the high energy of optical phonons in graphene quench their contribution to electron-phonon cooling in graphene. While cooling through substrate phonons can also occur,\cite{fratini,price12} here we will only consider cooling mediated through the exchange of acoustic phonons in graphene. 

While $\mathcal{H}_{\rm el-ph}$ is central to electronic cooling,
as we will see disorder plays a surprising role. 
The effect of the disorder potential can be modeled as a sum of impurity potentials and a gauge field originating from strain that couples to electron velocity,\cite{guinea, katsnelson, neto, katsnelson2}
\be\label{eq:H_impurity}
\mathcal{H}_{\rm dis}=  \sum_{ \vec r, \alpha} 
\psi_{\alpha}^\dag(\vec r)\lb \sum_i V(\vec r-\vec r_i)+v\sigma\cdot\vec A(\vec r)\rb \psi_{\alpha}(\vec r)
.
\ee
The second type of disorder is peculiar for graphene, where it can arise due to disorder-induced ripples on the graphene sheet.\cite{katsnelson, neto, katsnelson2, ishigami, folk}
The random vector potential $\vec A(\vec r)$
depends on the microscopic ripple profile.

We begin to analyze electronic cooling (middle panel Fig.\,\ref{pe}) by considering electron-phonon scattering rate described by Fermi's Golden Rule, 
\bea \label{transition}
W_{\vec{k}',\vec{k}}&=& \frac{2\pi}{\hbar}\sum_{\vec q}\lb |M_+|^2 N_{\omega_{\vec q}}\delta( \epsilon_{\vec{k}'}-\epsilon_{\vec{k}}- \omega_\vec{q})\right.
\\\nonumber
&&\left. +|M_-|^2 (N_{\omega_{\vec q}}+1)\delta( \epsilon_{\vec{k}'}-\epsilon_{\vec{k}}+ \omega_\vec{q})\rb
,
\eea
where $\vec{q}$ is phonon momentum, and $N_{\omega_{\vec q}} = 1/(e^{\beta \omega_\vec{q}}-1)$ is the Bose distribution.
In the absence of disorder, the matrix elements are derived from the deformation potential, $M_{\pm}^{(0)}=g\sqrt{\omega_{\vec q}}\delta_{\vec k'-\vec k\mp\vec q}$, where the delta function enforces momentum conservation. This gives the cooling power
\be
\mathcal{J}= \sum_{\vec k, \vec{k'}, i} W_{\vec{k}', \vec{k}} (\epsilon_\vec{k} - \epsilon_\vec{k'}) f(\epsilon_\vec{k}) \big[ 1- f(\epsilon_\vec{k'})\big]  
\label{cool}
\ee
where $f(\epsilon) = 1/(e^{\beta(\epsilon -\mu)} + 1)$ are Fermi functions, $W_{\vec{k}', \vec{k}}$ is the transition probability, and $\epsilon_\vec{k} - \epsilon_\vec{k'}$ is the energy exchanged in each scattering event.  Here we note that since thermalization (through carrier-carrier scattering, see Sec.\,\ref{sec:ie}) occurs 
%
far faster than the cooling time scales described here, a separation of time scales allows us to treat the electron distribution as a Fermi distribution with a higher temperature, $T_{\rm el}$, than the lattice, $T_{\rm ph}$.

\begin{figure}
\includegraphics[scale=0.5]{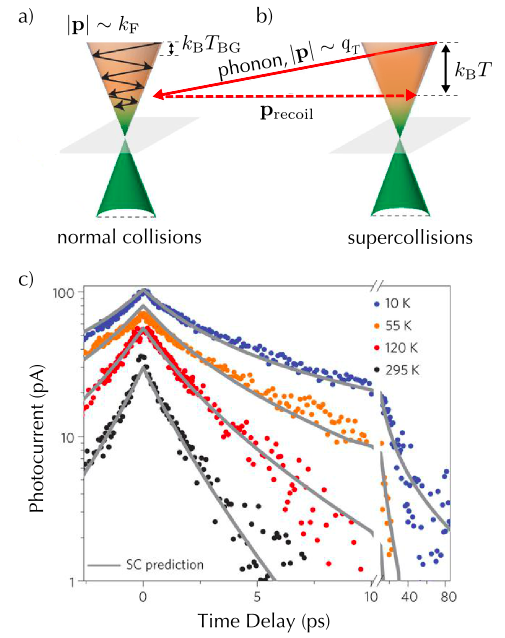}
\caption{ a,b) Kinematics of electron cooling in graphene showing (a) the normal momentum conserving process (takes many steps), and (b) the supercollision process (takes few steps). In (a), phonon momenta are constrained by the Fermi surface so that they carry off energy $\sim k_{\rm B} T_{\rm BG} \approx {\rm several}\, {\rm meV}$ (for typical carrier densities, see for example Ref. \onlinecite{graham,betz2013}) 
In (b), phonon momenta are totally unconstrained for supercollisions, with the recoil momentum ($|\vec p_{\rm recoil}| \approx$ $q_T$) transferred to the lattice via disorder scattering. 
The energy scales in this illustration have been exaggerated to illustrate the kinematics of phonon emission; hot electrons are thermal and close to the Fermi surface. c) Electron temperature dynamics extracted from a photocurrent experiment\cite{graham} (see text) showing two distinct regimes of cooling expected from supercollisions: i) an initial power law cooling dynamics (see Eq. \ref{eq:1/t}), and ii) later exponential regime (see Eq. \ref{eq:fulldynamics}); photocurrent is proportional to the electronic temperature. Gray lines denote predicted dynamics from supercollisions [Eq. (\ref{eq:fulldynamics})]. Illustration in (a,b) and experimental data (c) adapted from Ref. \onlinecite{graham}.}
\label{fig:sckinematics}
\end{figure}

The small Fermi surfaces in graphene allow easy access to two distinct regimes of electron-phonon scattering, namely $T>T_{\rm BG}$, and $T<T_{\rm BG}$. Here $T_{\rm BG} = s\hbar k_{\rm F}$ is the Bloch Gr\"uneissen temperature, which are tens of kelvin for typical densities in graphene; $k_{\rm F}$ is the Fermi wavevector. In the more familiar $T<T_{\rm BG}$ limit, electron-phonon scattering is constrained by the small available phonon phase space set by the low temperature;\cite{hwang08} this is analogous to electron-phonon scattering in metals.\cite{ziman} The Bloch-Gr\"uneissen regime exhibits suppressed cooling powers that decrease as temperature is decreased, giving $\mathcal{J} \propto (T_{\rm el}^\delta - T_{\rm ph}^\delta)$ where $\delta = 4$.\cite{kubakaddi,viljas} Combined with a small electronic heat capacity, $T<T_{\rm BG}$ feature long cooling times that vary between nanoseconds to microseconds in the several Kelvin to tens of milliKelvin range.\cite{kubakaddi,viljas,schwab,betzprl} The extremely small cooling powers achievable at ultracold temperatures allow for increased sensitivity for graphene bolometers.\cite{schwab}

Surprisingly, long cooling times from the emission of acoustic phonons can also occur in the regime $T> T_{\rm BG}$ that can be as long as a few nanoseconds.\cite{macdonald,wong} 
This appears because the mismatch of Fermi and sound velocities, $v/s \approx 100$, together with
the conservation of energy and momentum severely constrain the available phase space. As a result, only very long wavelength acoustic phonons (i.e. weakly coupled) are emitted.  
Indeed, the phonons exchanged have energy $\omega_\vec{q}  \approx s\hbar k_{\rm F} = k_{\rm B} T_{\rm BG} $, a small fraction of $k_{\rm B} T$ for practically interesting temperatures (see Fig.\,\ref{fig:sckinematics}a). This yields a cooling power that is suppressed \cite{macdonald,wong}
\be
\mathcal{J}_0 =  B ( T_{\rm el} - T_{\rm ph}), \quad B = \pi N \hbar g^2 [\nu (\mu)] ^2 k_{\rm F}^2 s^2 k_{\rm B}. 
\label{eq:pconserving}
\ee
where $N(\omega_{\vec q}) \approx k_{\rm B}T/\omega_{\vec q}$. As a result, cooling times can be as long as several nanoseconds even at room temperature, and the cooling rate exhibits an anomalous temperature dependence {\t increasing} with decreasing temperature.\cite{macdonald,wong}

The cooling power of electrons in $T> T_{\rm BG}$ can be dramatically enhanced by supercollision processes that can relieve the bottleneck from the momentum conserving normal process. This is most easily demonstrated by electron-phonon scattering in the presence of disorder (see Fig.\,\ref{fig:sckinematics}b for an illustration). Since disorder breaks translational invariance, emitted (absorbed) phonon momenta in Eq.(\ref{transition}) can take on unconstrained values, $|\vec q|\lesssim q_T$. As a result, thermal phonons are exchanged in supercollision electron-phonon scattering processes boosting the exchange of energy between electron and lattice systems.

At low disorder concentration, we can describe supercollision electron-phonon scattering by dressing the electron-phonon vertex with multiple scattering on a single impurity. 
This gives an expression for the transition matrix elements $M_\pm$ which is exact in the impurity potential:
\be
M_\pm = \la \vec k'|M_{\pm}^{(0)}G\hat T +\hat TGM_{\pm}^{(0)} +  \hat TGM_{\pm}^{(0)}G\hat T |\vec k \ra
,
\label{matrixelement_M}
\ee
where $G(\vec p)=\frac1{\epsilon-H_0(\vec p)+\mu}$ is the electron Green's function, $\hat T$ is the T-matrix (scattering operator) for a single impurity.
Scattering off an impurity relieves the momentum conservation that originally constrained the phase space for single acoustic phonon scattering allowing higher energy thermal phonons to be emitted.  
The supercollision transition matrix element $M_\pm$ are smaller than that of single acoustic phonon emission $M_\pm^{(0)}$ resulting in a smaller frequency of scattering.\cite{song12} However, the ability to emit thermal phonons more than makes up for the decreased frequency in scattering, thereby boosting cooling powers. Indeed, supercollisions dominate electron-lattice cooling over a wide range of technological relevant temperatures, see Table I.

In the following, we lay out a variety of supercollision scattering processes and calculate their cooling powers. We begin first with supercollision cooling that arises from scattering off short-range impurity scatterers which will serve as a conceptually simple illustration of supercollisions; other supercollision processes follow from similar reasoning (see Table I 
and discussion below) and yield qualitatively similar cooling powers.

\begin{widetext}
\begingroup
\squeezetable 
 \begin{ruledtabular}
\begin{tabular}{l c c r}
Supercollision Process & Cooling Power (Linearized) & Crossover Temperature, $T_*$ & Equation\\
\hline
Short-Range Disorder + 1 Acoustic Phonon & $ T^2 \Delta T$ & $\approx {\rm few }\, T_{\rm BG}$ &Eq.(\ref{eq:Jsuper})\\
Ripple Disorder + 1 Acoustic Phonon & $ T^2 \Delta T$ & $\approx {\rm few } \, T_{\rm BG}$ &Eq.(\ref{eq:Jripple}) \\
Coulomb Disorder + 1 Acoustic Phonon & vanishes & - &Eq.(\ref{eq:coulomb})\\
Two Acoustic Phonons: Anharmonic Coupling & $T^5 \Delta T$ & $\approx 20 \,T_{\rm BG}$ &Eq.(\ref{eq:2phononvertex})\\
Two Acoustic Phonons: Deep Off-Shell Process & vanishes at leading order & - &Eq.(\ref{eq:sequential2phonon})\\
Two Flexural Phonons & $T^2 \Delta T$ & $\approx 10 \,T_{\rm BG}$ &Eq.(\ref{eq:2F_cooling_power})\\
\end{tabular}
\end{ruledtabular}
\endgroup
\medskip
\noindent
Table I: Various supercollision processes in monolayer graphene and their linearized cooling power for $T> T_{\rm BG}$. $T_*$ indicates the crossover temperatures above which the supercollision process dominates over momentum conserving single acoustic phonon emission, $\mathcal{J}_0$.
\end{widetext}

\subsection{Short Range Disorder-Assisted Supercollisions}
Here we consider scattering by short-range impurities (the first term in Eq.(\ref{eq:H_impurity})) first discussed in Ref. \onlinecite{song12}. Short-range disorder can be modeled by a delta function potential 
\be 
V(\vec r-\vec r_j)=u\delta(\vec r-\vec r_j)(\hat 1\pm\sigma_z)/2,
\ee
where the plus (minus) sign corresponds to the impurity positions on the A (B) sites of the carbon lattice, and $u$ is the strength of the short-range scatterer. Focussing only on contributions first order in $u$, we can approximate $\hat T_{\vec p',\vec p}=\frac12 u (\hat 1\pm\sigma_z) e^{i(\vec p'-\vec p)\vec r_j}+O(u^2)$ in Eq.(\ref{matrixelement_M}). 

As we shall see, the main contribution to cooling will arise from phonons with momenta of order $q_T$. Thus we anticipate that the virtual electron states in the transition matrix element Eq.(\ref{matrixelement_M}), described by the Green's functions $G(\vec p)$, are characterized by large momenta $|\vec p|\sim q_T$ which are much greater than $\vec k$, $\vec k'$. In this case, for the off-mass-shell virtual states such that $v|\vec p|\gg \mu,\, k_{\rm B}T$, we can approximate $G(\vec p)\approx -\frac1{H_0(\vec p)}$. This form of $G(\vec p)$ describes deep off-shell virtual states, i.e. virtual states far from mass shell. The stiffness of the electron dispersion, $v\gg s$, along with the estimate $|\vec p|\sim q_T$, makes it an accurate approximation for all virtual states not too close to the Fermi surface. In this limit, as we now show, drastic simplifications occur because of particle-hole symmetry $H_0(-\vec p)=-H_0(\vec p)$.

Concentrating on contributions first order in $u$, we evaluate the first two terms in Eq.(\ref{matrixelement_M}). This gives the commutator of $H_0^{-1}(\vec q)$ and $\pm\sigma_z$  arising because the virtual electron states in the first and second term have momenta $\vec p\approx -\vec q$ and $\vec p\approx +\vec q$ (see above). We obtain
\be\label{eq:M1}
M_\pm = \frac{\pm i u g\sqrt{\omega_{\vec q}}}{\hbar v|\vec q|^2}\la \vec k'|(\sigma \times\vec q)_z |\vec k\ra e^{i(\vec p'-\vec p \pm \vec q)\vec r_j}
,
\ee
with the phase factor describing the dependence on the impurity position. We note that the finite commutator, $[\sigma_z, \sigma \cdot \vec q] = 2i (\hat{\vec z} \times \vec q) \cdot \sigma$ arises because of the matrix/sublattice structure of the impurity potential used. As we will see later, this structure is critical; for impurities that have an identity structure in pseudospin space, $M$ vanishes. In the following, we proceed to evaluate the energy-loss power in Eq.(\ref{cool}). 

Integrating Eq.(\ref{cool}) using Eq.(\ref{eq:M1}) in the same way as detailed above, and summing over A(B) impurities after squaring $|M|$,\cite{ando} the supercollision cooling power for $\mu \gg k_{\rm B}T$ is\cite{song12}
\be\label{eq:Jsuper}
\mathcal{J}_{\rm dis}= A_{\rm dis}\lp T_{\rm el}^3-T_{\rm ph}^3\rp
, \qquad 
A_{\rm dis}=9.62 \frac{g^2\nu^2(\mu) k_{\rm B}^3 }{\hbar k_{\rm F}\ell}
,
\ee
where the mean free path $k_{\rm F}\ell=2\hbar^2 v^2/( u^2 n_0)$.\cite{ando}

It is useful to compare the supercollision cooling power in Eq.(\ref{eq:Jsuper}) with the momentum conserving channel in Eq.(\ref{eq:pconserving}). The enhancement factor of disorder-assisted cooling over the momentum conserving pathways depends on both disorder and temperature:
\be
\frac{\mathcal{J}_{\rm dis}}{\mathcal{J}_0}=\frac{0.77}{k_{\rm F}\ell}\frac{T_{\rm el}^2+T_{\rm el}T_{\rm ph}+T_{\rm ph}^2}{ T_{\rm BG}^2}
,
\label{eq:enhancement}
\ee
where we used the value of $g$ described above. 
At room temperature, $T_{\rm el(ph)}\sim 300\,{\rm K}$, and taking $\mu =100\, {\rm meV}$ ($n \sim10^{12}\,{\rm cm^{-2}}$) we find $T_{\rm el(ph)}/T_{\rm BG}\approx 50$. 
For $k_{\rm F}\ell =20$,
the enhancement factor ${\mathcal{J}}/{\mathcal{J}_0}$ can be as large as 100 times.  For small $\Delta T = T_{\rm el} - T_{\rm ph}$, we find that $\mathcal{J}_{\rm dis}$ dominates over $\mathcal{J}_0 $ at temperatures
$
T>T_*=\sqrt{\frac{B}{3A}}=\lp \frac{\pi}{6\zeta(3)} k_{\rm F}\ell\rp^{1/2} T_{\rm BG}
.
$
Taking $k_{\rm F}\ell=20$ for a rough estimate, we see that the disorder assisted cooling channel dominates for $T\gtrsim 3T_{\rm BG}$. The crossover temperature  can be controlled by gate voltage, since $T_{\rm BG}\propto \sqrt{n}$. For typical carrier densities $n$ this gives a crossover temperature $T_*$ of a few tens of kelvin (see Table I).

Supercollision cooling in Eq.(\ref{eq:Jsuper}) is not only large, but it also exhibits qualitatively different features from the normal momentum conservation process that include, a cooling rate that decreases with decreasing temperature and two-stage cooling dynamics (discussed below), as well as a linear dependence on carrier density, see Eq.(\ref{eq:Jsuper}). In contrast, normal momentum-conserving electron-phonon processes (single acoustic phonon emission) depends quadratically on carrier density.

\subsection{Temperature Dynamics}

The temperature dynamics of hot carrier cooling can be described by accounting for both disorder-assisted supercollisions and momentum-conserving cooling, yielding hot carrier relaxation dynamics %
\be\label{eq:dQ/dt}
d\mathcal{Q}/dt=-\mathcal{J}_{\rm dis} -\mathcal{J}_0
,
\ee
where $\mathcal{Q}$ is the electron energy density. For small $\Delta T \ll T_{\rm el}, T_{\rm ph}$, and taking $\mathcal{Q}=C\Delta T$, with  $C=\alpha T$ the heat capacity of the degenerate electronic system ($\alpha=\frac{\pi^2}3N\nu(\mu)k_{\rm B}^2$), the cooling times describing relaxation to equilibrium, $\Delta T_{\rm el} (t) =   e^{- (t-t_0)/\tau}\Delta T_{\rm el,0}$, exhibit a nonmonotonic $T$ dependence:
\be
\label{eq:nonmono}
\frac{1}{\tau}=\frac{3A}{\alpha}T+\frac{B}{\alpha T}.
\ee
The cooling time increases with $T$ at $T<T_*$ and decreases at $T>T_*$, reaching maximal value at $T=T_*$. This non-monotonic temperature dependence in $\tau$ provides a clear experimental signature of the competition between different cooling pathways (see Sec.\,\ref{sec:exp}).

On the other hand for $T_{\rm el} \gg T_{\rm ph}$, the energy-relaxation power is $d\mathcal{Q}/dt \approx AT_{\rm el}^3$. This yields a $1/(t-t_0)$ dynamics:\cite{song12}
\be\label{eq:1/t}
T_{\rm el}(t)
=\frac{T_{\rm el,0}}{1+(A/\alpha)(t-t_0)T_{\rm el,0}}
.
\ee

For large initial electron temperatures $T_{\rm el} \gg T_{\rm ph}$, both the dynamics for $T_{\rm el} \gg T_{\rm ph}$ and $T_{\rm el} \approx T_{\rm ph}$ can be accessed at early and long times respectively, resulting in a distinct two-stage cooling dynamics. The full dynamics can be found by directly solving Eq.(\ref{eq:dQ/dt}), yielding \cite{song12}
\be
\label{eq:fulldynamics}
-\frac{2}{\tau} (t-t_0) = F(T_{\rm el}(t)/ T_{\rm ph}) - F(T_{\rm el,0}/T_{\rm ph}),
\ee  
where $F(x) = 2\sqrt{3} \arctan [(1+2x)/\sqrt{3}] - {\rm ln} [ (x^3-1)/(x-1)^3]$ ($\mathcal{J}_0$  has been suppressed only becoming important for $T\lesssim T_*$ and at long times). This solution is characterized by only two parameters: the initial dimensionless temperature $x_0 = T_{\rm el,0}/T_{\rm ph}$ and the time constant $\tau = 3 T_{\rm ph} A /\alpha$ allowing a clear way of comparing with experiments that probe the carrier dynamics of graphene.

\subsection{Experimental Observation of Supercollisions}
\label{sec:exp}

Observing the cooling mechanisms in graphene have been the subject of a number of recent experiments.\cite{rana,graham,grahamnano,betz2013,qiong,somphosane13,laitinen2014} Indeed by accessing various aspects of electronic cooling, these have found that supercollisions dominates electron-lattice energy relaxation in graphene over a wide range of temperatures. 
%
The methods employed can be broadly classified as i) dynamical measurements of temperature dynamics,\cite{rana,graham,grahamnano} and ii) steady-state measurements of cooling power.\cite{betz2013,qiong,somphosane13,laitinen2014} We briefly review a few of these experiments. 

Dynamical measurements of hot carrier temperature dynamics provide a powerful method in identifying cooling channels in graphene.\cite{rana,graham,grahamnano} 
Ref. \onlinecite{rana} tracked the cooling dynamics of hot carriers in an optical pump-THz probe experiment and found that cooling times increased as lattice temperature was decreased. This is consistent with supercollisions described above but contrasts with normal collisions where the opposite behavior is expected (see Eq. \ref{eq:pconserving}). 

In a similar vein, Ref. \onlinecite{graham} extracted the cooling dynamics of graphene under pulsed illumination in a pump-probe setup,\cite{graham} using photocurrent from the photothermoelectric effect as a thermometer of the electrons in graphene.\cite{graham,grahamnano,gabor} In these measurements, Ref. \onlinecite{graham} found a distinct two-regime cooling dynamics as shown in Fig.\,\ref{fig:sckinematics}c. When pumped intensely with pulsed irradiation, the electronic temperature decayed via an initial power law regime, and a later exponential regime matching the predictions of Eq. (\ref{eq:1/t}), (\ref{eq:fulldynamics}) above (see theoretical prediction in gray line coinciding with experimental data in Fig.\,\ref{fig:sckinematics}c). Further by controlling the initial excited electron temperature, Ref. \onlinecite{graham} was able to vary the power-law regime time constant confirming the $T_{\rm el,0}$ scaling predicted in Eq. (\ref{eq:1/t}).



\begin{figure}
\includegraphics[scale=0.33]{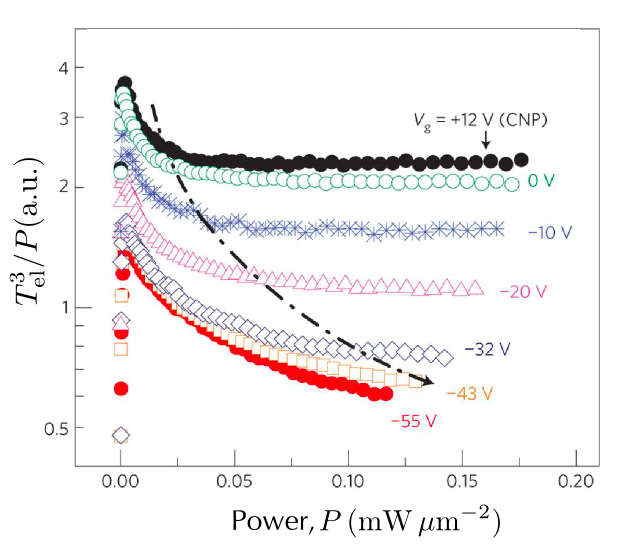}
\caption{Experimental observation of supercollisions (see also Fig.\,\ref{fig:sckinematics}c for a dynamical measurement). Steady state electronic temperature follows a cubic power law measured through noise thermometry in Ref. \onlinecite{betz2013}, as shown by the flat plateau regions [see text, and $T_{\rm el} \gg T_{\rm ph}$ regime in Eq. (\ref{eq:Jsuper})]. Note that the crossover temperature to a supercollision dominated regime (plateau) is controlled by gate potential via the density dependence of $T_{\rm BG} = s\hbar k_{\rm F}$ [see Eq. (\ref{eq:enhancement})].}
\label{fig:exp}
\end{figure}

Ref. \onlinecite{betz2013} adopted an alternative steady state approach in which they measured electronic temperature through noise thermometry. By Joule heating the electronic system, and subsequently extracting its (steady state) electronic temperature, Ref. \onlinecite{betz2013} were able to establish that for lattice temperatures larger than $T_{\rm BG}$, the electronic temperature scaled as $T_{\rm el} \propto P^{1/3}$ in agreement with Eq. (\ref{eq:Jsuper}) (plateau in Fig.\,\ref{fig:exp}); here $P$ is the power pumped into the system. Ref. \onlinecite{graham} also found the same scaling when they used a continuous wave illumination. Further, they varied the onset of the plateau region via gate voltage; the smaller the size of the Fermi surface the earlier the onset of supercollisions [see Eq. (\ref{eq:enhancement})]. Additionally, they extracted a linear density dependence of the electron-lattice cooling power [see $A_{\rm dis} \propto [\nu(\mu)]^2$ in Eq. (\ref{eq:Jsuper})]; this contrasts with the quadratic density dependence expected from normal collisions [Eq. (\ref{eq:pconserving})].

Another interesting aspect of electronic cooling in graphene is the non-monotonic temperature dependence of cooling times, Eq. (\ref{eq:nonmono}), that arises from the competition between normal process (dominating for $T< T_*$) and supercollisions (dominating for $T>T_*$). In Ref. \onlinecite{qiong}, a steady state photothermoelectric photocurrent technique was employed that tracked both the electronic temperature and cooling lengths in high quality gated graphene devices. They observed a non-monotonic temperature dependence of cooling times [Eq. (\ref{eq:nonmono})] peaking at an intermediate temperature $T_*$. Further, Ref. \onlinecite{qiong} demonstrated control of the peak temperature $T_*$ via gate voltage and disorder (through an annealing technique).\cite{qiong}

\subsection{Coulomb Impurities}

A large variety of disorder types (short-range, long-range Coulomb scatterers, ripples) can exist in graphene; it is important (on both the technological and fundamental levels) to consider what kind of disorder is necessary for supercollisions. An important type of disorder are Coulomb scatterers, which are believed to dominate the electronic transport characteristics of graphene.\cite{dassarmareview}

To analyze the effect of Coulomb impurities, we proceed in the same way as detailed above with the transition matrix element for supercollision scattering described by Eq.(\ref{matrixelement_M}). Using the Born approximation, we write $\hat {T}_{\vec p',\vec p}=\frac{2\pi Ze^2}{\kappa |\vec p'-\vec p|}$, where $Z$ is the impurity charge and $\kappa$ is the dielectric constant. We take the unscreened Coulomb potential since the main contribution to cooling come from phonons with momenta $q_T \gg k_{\rm F}$. Additionally, since Coulomb scatterers are long-ranged, they do not differentiate between sublattices; any differences arising from having Coulomb impurities on different sites can be absorbed into the scattering from short-range scatterers (see above). 

To evaluate the matrix element, we begin with the first two terms of Eq.(\ref{matrixelement_M}). Following the analysis above, the sum of the first two terms of Eq.(\ref{matrixelement_M}) evaluates approximately as the commutator between $G(\vec p)$ and $\hat{T}$. Since $\hat{T}$ for Coulomb scatterers is an identity matrix in graphene's pseudospin space, this commutator {\it vanishes} and the sum of the first two terms evaluates to zero. This cancellation originates from the equal (but opposite in sign) contributions of positive and negative energy virtual states. The physical origin of such ``destructive interference effect'' can be linked to the particle-hole symmetry of graphene's electronic spectrum, $H_0(-\vec p)=-H_0(\vec p)$ (see Sec.\,\ref{sec:deepoffshell} for a full discussion).

We proceed to consider the last term in Eq.(\ref{matrixelement_M}). Noting that the dominant contributors to cooling arise from phonons with momenta $q_T$ we estimate 
\be\label{eq:M2}
M_\pm = \frac{g\sqrt{\omega_{\vec q}}}{\kappa^2}
\sum_{\vec p}\la \vec k'|\frac{(2\pi Ze^2)^2}{|\vec p-\vec q||\vec p|}H_0^{-1}(\vec p-\vec q)H_0^{-1}(\vec p)|\vec k\ra
.
\ee
We can evaluate this by rewriting $H_0^{-1}(\vec p)/ |\vec p| = \sigma \cdot \nabla_\vec{p} \frac{1}{|\vec p|}$. Performing a 2D Fourier transform (to the conjugate variable $\vec{r}$) we obtain
\be
M_\pm = c \int d^2 \vec{r} \la \vec k'| \frac{(\sigma \cdot \vec r)^2}{|\vec r|^2 } e^{i \vec q \cdot \vec r} |\vec k\ra =  c  (2\pi)^2  \delta( \vec{q}) \la \vec k' |\vec k\ra,
\label{eq:coulomb}
\ee
where $c = g \sqrt{\omega_\vec{q}} ( 2\pi Z e^2)^2 / ( \kappa^2 v^2 \hbar^2)$. Since the contributions we are interested in are $|\vec q| \gg k_{\rm F}$, the matrix element {\it vanishes}. Hence, we conclude that Coulomb scatterers do not contribute to the supercollision cooling channel. This observation means that, while Coulomb scatterers are central for the electronic transport characteristics of graphene, they contribute very little to the energy relaxation pathways of hot carriers. Coulomb scatterers are therefore not as important for graphene  optoelectronic properties as they are for transport properties.

\subsection{Ripple-assisted Supercollisions}


Another type of disorder arises from ripples in graphene. Ripples can originate in a number of ways. These include: (i) {\it intrinsic} ripples which occur in pristine and suspended graphene, and (ii) {\it extrinsic} undulations caused by corrugations of an underlying substrate. Both of them cause
electrons to be scattered off random strain in the graphene sheet. 
This type of disorder is described by the vector potential term in Eq.(\ref{eq:H_impurity}).  
Evaluating the electron-phonon matrix element, Eq.(\ref{matrixelement_M}), we find a nonzero contribution at lowest order in $\vec A_{\vec q}=\int d^2r e^{-i\vec q\vec r}\vec A(\vec r)$. The estimate proceeds in the same way as for short range disorder, with $G(\epsilon,\vec q) \approx 1/ H_0(\vec q)$ describing deep off-shell states. The matrix structure $\sigma\cdot\vec A(\vec r)$ generates the commutator $[\sigma\cdot\vec A_{\vec q}, H_0^{-1}(\vec q)]=\frac{-1}{\hbar v|\vec q|}2i\sigma_z \tilde{\vec q}\cdot\vec A_{\vec q}$, giving 
\be 
M_\pm(\vec A_\vec{\vec q})= \la \vec k'|\frac{2 g\sqrt{\omega_q} i \sigma_z }{\hbar  |\vec q| } \tilde{\vec q}\cdot \vec A_{\vec q} | \vec k\ra
.  
\label{matrixele3}
\ee
Evaluating the transition rate $W_{\vec k',\vec k}$ and plugging it into Eq.(\ref{cool}), we obtain  the cooling power
\be
\mathcal{J}_{\rm ripple}= b 
\sum_{\vec q}\omega_q \big( \tilde N_{\omega_{\vec q}} - N_{\omega_{\vec q}} \big) \la | \tilde{\vec q}\cdot \vec A_{\vec q} |^2\ra_{\rm dis}
,
\label{eq:J_ripples}
\ee
where $b= 4\pi N g^2 s^2 \big[\nu(\mu) \big]^2/\hbar$ and averaging over the ripple ensemble is denoted by $\la ...\ra_{\rm dis}$. 

We note that the pair correlator $\chi(\vec q)=\la A_\vec{q} A_{-\vec q} \ra_{\rm dis}$ decays at $|\vec q|\approx q_0= 2\pi/R$, where $R$ is the characteristic radius of a ripple.\cite{katsnelson} Here we assume that the radius $R$ is much shorter than the phonon wavelength for temperatures of interest (indeed, Ref. \onlinecite{folk} estimates $R = 4  \, {\rm nm}$, whereas at room temperature $\lambda_T=2\pi/q_T \approx 17 \, {\rm nm}$).  
Since at $q_T\ll q_0$ the integral over $\vec q$ in the general expression (\ref{eq:J_ripples}) is limited solely by the Bose functions $N_{\omega_{\vec q}},  \tilde N_{\omega_{\vec q}}$,  the correlator $\chi(q)$ is approximately given by its value at $|\vec{q}|\ll q_0$. This allows us to evaluate the integral over the Bose functions treating $\chi(q)$ as a constant. The correlator $\chi(\vec q)$ was analyzed in Ref. \onlinecite{katsnelson}.
For the roughness exponent $2H \approx 1$ corresponding to graphene on a substrate,\cite{ishigami} 
Ref. \onlinecite{katsnelson} obtains 
$\chi(|\vec q|\ll q_0)\approx (\hbar/a)^2 Z^4/R^2$ 
where $Z$ is the characteristic out of plane displacement in the graphene sheet caused by ripples and $a$ is the interatomic distance. This yields a cooling power 
\be\label{eq:Jripple}
\mathcal{J}_{\rm ripple}= A_{\rm ripple}\lp T_{\rm el}^3-T_{\rm ph}^3\rp
,
\ee
where $A_{\rm ripple}$ is 
\be
A_{\rm ripple}= \frac{2 \zeta(3) k_{\rm B}^3 b}{2\pi\hbar^2 s^2} \int \frac{d\theta_q}{2\pi} \la |  \tilde{\vec q} \cdot \vec A_{|\vec{q}|\ll q_0}|^2\ra_{\rm dis}
= \frac{\zeta(3)b Z^4 k_{\rm B}^3}{ 2\pi s^2 a^2 R^2}
,
\ee
where we accounted for a factor of $1/2$ due to the angular integral.
Comparing this result with the contribution of normal collisions in Eq.(\ref{eq:pconserving}),
we obtain a crossover temperature
\be\label{eq:ripplecomp}
T_*=\eta^{-1/2}T_{\rm BG}
,\quad
 \eta = \frac{6 \zeta(3)}{\pi} \frac{Z^4}{R^2a^2}
.
\ee
Taking the values reported in Ref. \onlinecite{folk}, $Z= 0.6 \, {\rm nm}$ and $R =  4 \, {\rm nm}$, we find $\eta \approx 0.9$, which gives $T_* \approx  T_{\rm BG}$. 
For a doping level of $\mu = 100 \, {\rm meV}$ this is $T_* = 12 \, {\rm K}$. For weaker disorder, we anticipate $T_*$ to be a few times $T_{\rm BG}$. Recently, shot-noise thermometry measurements in Ref. \onlinecite{laitinen2014} measured cooling in suspended graphene samples and attributed it to ripple-assisted supercollisions.

\subsection{Two-Acoustic Phonon Emission: Anharmonic Coupling}
\label{sec:anharmonic}

Next, we consider two-phonon scattering which can utilize the supercollision phase space enhancement in much the same way as the disorder-assisted channel described above; two-phonon processes considered in this section (and Sec.\,\ref{sec:deepoffshell}, \ref{sec:flexural}) are intrinsic processes which will also occur in pristine graphene (in the absence of disorder). This process is described by an anharmonic coupling\cite{ABO, marzari} 
\be
\mathcal{H}_{\rm el-2ph}=D'\sum_{\vec{q_1},\vec{q_2}} n_{-\vec {q}_1-\vec {q}_2} u_{\vec{q_1}} u_\vec{q_2} 
,
\ee
where $u_\vec{q} = \sqrt{\omega_\vec{q}/2\rho s^2 }(b_\vec{q} + b^\dag_{-\vec q})$, and $D'$ is an energy of order unity in atomic units. 
The terms of the form $b^\dag_{\vec{q}_1} b_{\vec{q}_2}$ describe Compton scattering of phonons by electrons. These processes have a negligible contribution to the cooling power and can be neglected.

\begin{figure}
\includegraphics[scale=0.32]{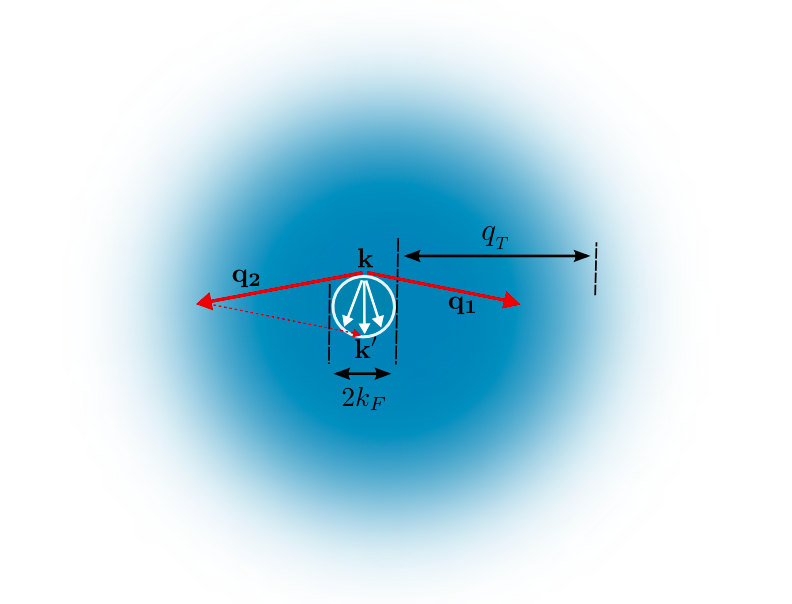}
\caption{Kinematics of two-phonon supercollisions wherein two nearly counterpropagating phonons, $\vec q_1$ and $\vec q_2$, are emitted. Here $\vec k, \vec k'$ denote the initial and final momenta of the electron, the blurred blue circle denotes the thermal distribution of phonons, and the white circle denotes the Fermi surface. The typical ratio of blue and white circle sizes is large $q_T/k_{\rm F} = v k_{\rm B}T/(s\mu) \approx 25$ (taking $T = 300 \, {\rm K}$, and typical $\mu = 100\, {\rm meV}$) making $\vec q_1 \approx - \vec q_2$.  
Phonons can be acoustic phonons (see Sec.\,\ref{sec:anharmonic},\ref{sec:deepoffshell}) or flexural phonons (see Sec.\,\ref{sec:flexural}).}
\label{fig:2phkinematics}
\end{figure}

Since $\vec k$, $\vec k'$ are on the Fermi surface, and the phonon momenta are large, $q_T \gg k_{\rm F}$, the two phonons {\it are nearly counterpropagating} (see kinematics in Fig.\,\ref{fig:2phkinematics}). We can thus set $\vec{q_1} \approx - \vec{q_2} = \vec{q}$. This simplifies the matrix element $M = D' \omega_{q}  / (2\rho s^2) $ yielding a transition probability
\be
W_{\vec k', \vec k} =  d \sum_\vec{q} |\la \vec{k'}|\vec k\ra|^2  \omega_\vec{q}^2\Big[ N_{\omega_{\vec q}}^2  \delta_+ + (N_{\omega_{\vec q}}+1)^2 \delta_-\Big]
,
\ee
where $d=\frac{2\pi N D'^2}{\hbar(2\rho s^2)^2}$, $\delta_\pm = \delta( \epsilon_{\vec k}-\epsilon_{\vec k'}\pm 2\omega_\vec{q})$. Here the coherence factor is $|\la \vec{k'}|\vec k\ra|^2  = \big[1 \pm {\rm cos} ( \theta_\vec{k} - \theta_\vec{k'} )\big]/2$, with the plus (minus) sign for intra-band (inter-band) processes. 

The two-phonon cooling power can be obtained from the transition probability by using Eq.(\ref{cool}). We find
\be
\mathcal{J}_{\rm 2ph}=\big[\nu(\mu) \big]^2 d \int \frac{d^2q}{(2\pi)^2} 4\omega_q^4 
{\cal G}(T_{\rm el}, T_{\rm ph}), 
\label{eq:2phononvertex}
\ee
where we defined
\be
{\cal G}(T_{\rm el}, T_{\rm ph}) =(N_{\omega_{\vec q}} + 1)^2 \tilde N_{2\omega_{\vec q}} -N_{\omega_{\vec q}}^2 ( \tilde N_{2\omega_{\vec q}} +1)
.
\ee
We note that 
${\cal G}$ vanishes when $T_{\rm el} = T_{\rm ph}$, satisfying detailed balance.  Expanding in small $\Delta T = T_{\rm el} - T_{\rm ph}$ and integrating over $\vec q$ we obtain the ratio $\mathcal{J}_{\rm 2ph}/\mathcal{J}_0 =  (D'/D)^2(T/T_0)^5$, where $k_{\rm B}T_0 = 0.86\,(\hbar^2 \rho s^4k_{\rm B}^2 T_{\rm BG}^2 )^{1/5}$.
Taking $D'$ of order unity in atomic units, $D'\sim e^2/a \sim 10 \, {\rm eV}$, we obtain a crossover temperature which for typical dopings is $5$-$10$ times larger than our $T_*$ estimate for disorder assisted processes (see Table I).
This means that the two-phonon processes can dominate only in very clean systems. The contribution of multi-phonon processes to energy transport has been considered in Ref.[ \onlinecite{virtanen}].

\subsection{Two Phonon Emission: Deep Off-Shell Processes}
\label{sec:deepoffshell}

In this section we consider a two-phonon process in which two counterpropagating phonons are emitted (or absorbed) through a virtual state (``off shell'' process). 
Each electron-phonon scattering vertex in this 
two-phonon process is described by  the standard deformation coupling, see Eq.(\ref{eq:H_elph}). 
As we now show, analyzing the transition amplitude for this process yields two deep off-shell processes that cancel at leading order.
We can see this by examining the matrix element for this process, 
\be
M = \la \vec{k'} |M^{(0)}_{\vec{q_1}} G(\epsilon,\vec q_2) M^{(0)}_{\vec{q_2}}+ M^{(0)}_{\vec{q_2}} G(\epsilon,\vec q_1) M^{(0)}_{\vec{q_1}}| \vec{k} \ra
\label{eq:sequential2phonon}
.
\ee
Here 
$
G(\epsilon,\vec q_1) = \frac1{\epsilon - H_0(\vec k-\vec{q_1}) +\mu}
$,
$
G(\epsilon,\vec q_2) = \frac1{\epsilon - H_0(\vec k-\vec{q_2})+\mu}
$
where $\mu$ is Fermi energy. 
Since $|\vec{q_1}|,|\vec{q_2}| \approx q_T \gg k_{\rm F}$, we can neglect $\vec{k},\vec{k'}$ and the two phonons scattered counterpropagate, i.e. $\vec{q_1} \approx - \vec{q_2} = \vec q$ (see Fig. \ref{fig:2phkinematics}).

It is instructive to consider the undoped system first, $\mu=0$. In this case, ignoring $\epsilon\sim kT$ compared to $H(q)\sim vq_T$, we have $G^{q_1} \approx -1/H_0(\vec q)$ and $ G^{q_2} \approx -1/H_0(-\vec q)$. We also note that the vertex $M^{(0)}$ is an identity in graphene's pseudospin space and it commutes with $H_0$. 
Since $H_0(-\vec q) = - H_0(\vec{q})$, the matrix element and the resulting transition probability vanishes. This, in turn, means that the leading contribution to cooling power for this process vanishes. 

Importantly, the above argument is not limited to the undoped case, since $v q_T\gg\mu$ for typical carrier densities and temperatures. Hence, even when the system is doped away from neutrality, the off-shell contribution is still negligible, see discussion below. 
On a conceptual level, the cancellation between the two terms in Eq.(\ref{eq:sequential2phonon}) may appear counterintuitive since the positive and negative contribution arise from the states above and below the Fermi level. We note in that regard, that the cancellation is made possible because no occupation factors are associated with $G(\epsilon,\vec q_1)$ and $G(\epsilon,\vec q_2)$ since they describe deep off-shell states as above. As a result, the nearly counterpropagating phonons allow the positive and negative energy contributions in the virtual states to cancel each other viz. particle-hole symmetry. 

In a similar vein, we can consider Compton-like processes in which phonons are absorbed and then re-emitted. The transition amplitude for such a process, similar to our discussion of two-phonon emission and absorption, is given by a sum of two contributions, one describing phonon absorption followed by phonon emission and the other describing phonon emission followed by phonon absorption. In contrast with the two-phonon emission and absorption discussed above,  phonon Compton scattering is dominated by processes in which the absorbed and emitted phonon have momenta that co-propagate. In complete analogy with our discussion above, the  symmetry $H_0(-\vec q) = - H_0(\vec{q})$ results in the transition matrix element vanishing at leading order. Interference between the term describing absorption and subsequent emission of a phonon and the term describing emission and subsequent absorption of a phonon resembles the QED treatment of Compton scattering. 

While the 
leading order contributions vanish for these processes, in each case there exists a residual contribution. Indeed, restoring finite $\epsilon$ to the Greens functions gives a non-vanishing sum 
\be
G(\epsilon,\vec q_1)  + G(\epsilon,\vec q_2) \approx \frac{2(\epsilon+\mu)}{[\epsilon - H_0(\vec{q}) +\mu][\epsilon + H_0(\vec{q})+\mu]},
\label{eq:sumofGs}
\ee
where we have imposed $\vec{q_1} \approx - \vec{q_2} = \vec q$ for counterpropagating phonons as before. 

We can follow the analysis used above to evaluate the cooling power for this residual contribution. In doing so, we note that the main contributions to supercollision cooling arise from momenta $ |\vec q| \approx q_T$. As a result, $\epsilon, \mu \ll v q_T$ in the degenerate limit ($\mu \gg k_{\rm B}T$) allows us to simplify the matrix element so that  $G(\epsilon,\vec q_1)  + G(\epsilon,\vec q_2) \approx - 2 \mu / [vq_T]^2$ in Eq.(\ref{eq:sumofGs}). This yields an estimate for the residual cooling power as
\be
\mathcal{J}_{\rm 2ph}^{\rm residual} = C\Big( T_{\rm el}^2 - T_{\rm ph}^2\Big), \quad C = \frac{N [\nu(\mu)]^2 \pi^2 s^2 g^4 \mu^2 k_{\rm B}^2}{3 \hbar v^4}. 
\ee
Interestingly, $\mathcal{J}_{\rm 2ph}^{\rm residual}$ scales the same way with density as single acoustic phonon cooling, $\mathcal{J}_0$. Comparing $\mathcal{J}_{\rm 2ph}^{\rm residual}$ above to $\mathcal{J}_0$ in Eq.(\ref{eq:pconserving}), by linearizing in small $\Delta T$, we estimate that $\mathcal{J}_{\rm 2ph}^{\rm residual} $ wins over $\mathcal{J}_0$ for $T > T_{*} \approx 5500 \, {\rm K}$. Here, we have used values for the electron-phonon coupling, $g$, detailed above. We therefore conclude that the residual contribution is irrelevant. 

\subsection{Flexural Phonon Supercollisions}
\label{sec:flexural}

Finally, we analyze cooling in free standing graphene in the absence of disorder. In this case, an important contribution arises due to flexural phonons,\cite{castro,ochoa} which contribute to the deformation tensor via $u_{ij} = 1/2 (\partial_i u_j + \partial_j u_i + \partial_i h\partial_j h)$, with $\vec{u}$ and $h$ the in-plane and out-of-plane displacements. Flexural modes have quadratic dispersion $\tilde\omega_\vec{q} =  \kappa_{\rm flex} |\vec q|^2$ with
$\kappa_{\rm flex}\approx 4.6 \times 10^{-7} \, {\rm m}^2 {\rm s}^{-1}$.\cite{castro,ochoa} Electron-phonon coupling is described by the same deformation potential as above, Eq.(\ref{eq:H_elph}).

The processes involving pairs of nearly counterpropagating flexural phonons are analyzed as follows.
Using the momentum representation, $h_\vec{q} = \sqrt{\hbar/2\rho \tilde\omega_q }\big(b_\vec{q} + b_{-\vec q}^\dag\big)$, 
we consider the emission/absorption of two flexural phonons with momenta $\vec{q_1}$ and $\vec{q_2}$. For $T \gg T_{\rm BG}^{\rm flex} = \hbar \kappa_{\rm flex} k_{\rm F}^2$ (for typical densities, $T_{\rm BG}^{\rm flex}$ is well below $1\,{\rm K}$), we can set $\vec{q_1} \approx - \vec{q_2} = \vec{q}$ so that the matrix elements in a transition rate are $M = \frac{D\hbar}{4 \rho \kappa_{\rm flex}}$.\cite{castro,ochoa} Integrating to obtain the cooling power yields \cite{song12} 
\[
\mathcal{J}_{\rm flex} = \sum_q
(2\hbar\tilde\omega_q)^2  \lb (N_{\omega_{\vec q}} \!\!+ 1)^2 N^{\rm el}_{2\omega_{\vec q}} -N_{\omega_{\vec q}}^2 ( N^{\rm el}_{2\omega_{\vec q}} \!\!+1)\rb
\]
where $\sum_q...=\frac{\pi N D^2\hbar^2 }{16 \rho^2 \kappa_{\rm flex}^2} \big[\nu(\mu) \big]^2 \int\!\! \frac{d^2q}{(2\pi)^2}...$, and the expression vanishes for $T_{\rm el} = T_{\rm ph}$.

For easy comparison, $\mathcal{J}_{\rm flex}$ can be linearized in $\Delta T = T_{\rm el} - T_{\rm ph}$ giving
\be\label{eq:2F_cooling_power}
\mathcal{J}_{\rm flex} = A_{\rm flex}  T^2 \Delta T, \qquad A_{\rm flex}  = 0.12\frac{ N D^2 \nu^2(\mu)k_{\rm B}^3}{\rho^2 \kappa_{\rm flex}^3},
\ee
which scales with $T$ the same way as Eq.(\ref{eq:Jsuper}). 
Flexural phonons dominate over the one-phonon contribution at
\be
T > T_*^{\rm flex}=\sqrt{\frac{B}{A_{\rm flex} }}=\Big( \frac{\pi  \rho \kappa_{\rm flex}^3}{0.24 \hbar s^2}   \Big)^{1/2} T_{\rm BG}  \approx 10 \,T_{\rm BG}
.
\ee
The value $T_*^{\rm flex}$ is considerably larger than $T_*$ for disorder-assisted cooling estimated above (see Table I).
 A comparison with Eq. (\ref{eq:Jsuper}) yields $\mathcal{J}_{\rm flex}/ \mathcal{J}_{\rm dis} \approx k_{\rm F} \ell/ 200$ which is small for typical $k_{\rm F}\ell$. Thus the contribution Eq. (\ref{eq:2F_cooling_power}) is relatively weak under realistic conditions.
For graphene on substrate this contribution is further diminished as flexural modes get pinned, gapped, and stiffened by the substrate.

\section{Conclusion}

As we have seen above, graphene  provides a natural setting to realize and explore long-lived hot carriers. Arguably, hot carriers is one of the most exciting and promising new developments in graphene research that relies on the  unique properties of this material. 
However, on a practical implementation level, it remains to be seen whether graphene hot carriers are 
 a pack of dogs that can bark.
This will happen as more new phenomena triggered by the presence of hot carriers are uncovered. 
Hot carriers have been proposed as a means to enhance optoelectronic efficiency,\cite{song,klaas,nozik} and manipulate energy flows in new ways.\cite{edrag,folk14} 
One can envision several other avenues where hot carrier research may leave a mark. We anticipate new transport phenomena mediated by long-range energy currents along the lines of Refs.\onlinecite{folk14,edrag}, in particular new kinds of photoresponse. Another interesting direction is energy harvesting in 2D crystals, such as graphene. One approach which is now actively discussed is vertical extraction of photogenerated hot carriers in field-effect or sandwich-type structures.\cite{britnell1} It is also of interest to explore various knobs by which the key properties of hot carriers such as cooling times can be altered and controlled.\cite{song12} Given the wide range of lattice temperatures for which hot carriers can be realized, extending up to room temperature, we have many reasons to expect that this research will eventually transcend into the practical realm.

\begin{acknowledgments}

We are grateful to F. H. L. Koppens, M. Y. Reizer and K. J. Tielrooij for collaboration on some of the topics discussed above and many useful discussions. We also thank A. J. Frenzel for useful discussions. This work was supported as part of the Center for Excitonics, an Energy Frontier Research Center funded by the U.S. Department of Energy, Office of Science, Basic Energy Sciences under Award No. DE-SC0001088 (photoexcitation cascade), and by the Office of Naval Research Grant No. N00014-09-1-0724 (cooling mechanisms). JS acknowledges financial support from the NSS program (Singapore).

\end{acknowledgments}


\begin{thebibliography}{99}
\bibitem{note} We use this opportunity to correct a factor of $\sqrt{2}$ in the denominator of $g_0$ in Eq. 16 of Ref. \onlinecite{ie}. Accounting for the new coupling constant, the estimate of the scattering rate yields $\tau_0 \approx 350 \, {\rm fs}$ (see also Eq. \ref{eq:estimateph} in this text).
\end{thebibliography}
\end{document}